\documentstyle [12pt]{article}
\evensidemargin\oddsidemargin
\begin{document}
\count0 = 1
\begin{titlepage}
\vspace{30mm}
 \title{\small{ASYMPTOTIC STATE VECTOR COLLAPSE\\
AND QED NONEQUIVALENT
REPRESENTATIONS  \\ }}
\author{S.N.Mayburov \thanks{E-mail ~~ Mayburov@SGI.LPI.MSK.SU ~~
 }\\
Lebedev Inst. of Physics\\
Leninsky Prospect 53\\
Moscow, Russia, 117924\\
\\}
\date{ 20 may 1997}
\maketitle
\begin{abstract}
 The state vector evolution in the interaction of measured pure state
 with the collective quantum system or the field 
  is analyzed in a nonperturbative QED formalism. As
 the model example the measurement of the electron final state scattered on
 nucleus or neutrino is considered. The produced
 electromagnetic bremsstrahlung    
contains the unrestricted number of soft photons resulting in
 the total radiation flux
 becoming the classical observable, which means the state vector collapse. 
 The evolution from the initial to the final system state is nonunitary and
  formally irreversible in the limit of the infinite time.
\end{abstract}
\vspace{20mm}
\vspace {20mm}
\small{   Submitted to Int. Journ. Theor. Phys.}
\end{titlepage}
\section { Introduction}
 The problem of the state vector collapse description in Quantum
Mechanics (QM) is still open despite the multitude of the proposed
models and hypothesis (D'Espagnat, \\
1990). 
 This paper analyses some microscopic dynamical models of the collapse
- i.e. the models which attempt to describe
 the interaction and the joint evolution of the measured state (particle)
 and the measuring device D (detector ) from the first QM principles. 
  Currently the most popular  of them are the different variants of  
 Decoherence Model which take into account also
the interaction of  the environment E with very large
 number  degrees of freedom (NDF) and D with small NDF (Zurek,1982).
 Yet this model meets the serious conceptual difficulties 
  resumed in so called Environment Observables Paradox (EOP)
  (D'Espagnat,1990).
  It  demonstrates that for any decoherence process at any time moment 
 at least  one observable $\hat{B}$ exists  which
 expectation value coincides with the value for the pure state
and differ largely
from the predicted for the collapsed mixed state.
Moreover it follows that in principle it's possible to restore the
system initial state which contradicts with the irreversibility
 expected for the collapse. In general EOP can be regarded
as an important criteria of the measurement models correctness.  

 Meanwhile it was proposed  that due to the very large internal NDF
 of the real macroscopic detectors  the problem 
 can be resolved by the methods of  nonperturbative Quantum Field 
 Theory (QFT) , which study the dynamics of the systems with    
  the infinite NDF (Neeman,1985),(Fukuda,1987).
The states manifold
of such systems is described by Unitarily Nonequivalent (UN) representations, 
which permit to resolve EOP as will be demonstrated below.

The main difficulty of this approach is that it can be 
correctly applied only for the measurements on the 
systems with not simply very large, but exactly 
infinite NDF. Meanwhile  the practical measuring
devices must have the finite mass and energy.     
 Here we consider QED bremsstrahlung  
   model of the collapse which satisfy  to all
 this demands simultaneously and without contradictions. 
 It evidences also that the collapse-like processes
 can occur not only in the macroscopic objects, but also on the
fundamental level of the elementary particles and fields.   

 Nonperturbative methods in QED were applied successfully 
for the study of the photon bremsstrahlung produced in any processes
of the charged particles  scattering on some target.
The total number of produced photons 
 with the energy larger than $k_0$,  is
  proportional to $e^2ln\frac{P_e}{k_0}$ ,i.e. it  grows
 unrestrictedly as $k_0\rightarrow 0$.. The perturbative Feynman diagram
methods by definition works only for the processes which probability
is small , while in this case it approximates to 1 (Itzykson,1980). 
 The nonperturbative formalism was developed initially for the semiclassical
 case when the charges movement is prescribed (classical)
 and the back-reaction of 
 the radiated electromagnetic (e-m) field  $\hat{A}_{\mu}(x)$ on the charge
 movement can be neglected - BRF condition (Friedrichs,1953).
 Consequently in this case electromagnetic current
 $J_{\mu}(x)$ is not the operator ,but c-value and 
 for the single electron scattering  its 4-dimensional
 Fourier transform is equal to:
\begin{equation}  
    J_{\mu}(k)=ie(\frac{p_{\mu}}{pk}-\frac{p'_{\mu}}{p'k})   
\end{equation} 
 where $p,p'$ are the  initial and final $e$ 4-momentum. In this case
BRF condition 
means that the sum of radiated photons momentum $|\vec{k}_s|$ is much
less then the electron momentum transfer in the scattering
 $|\vec{p}-\vec{p'}|$ (Akhiezer,1981).  

Final e-m filed state is found  by the nonperturbative calculation
of S-matrix (S-operator) -  T-product of interaction
 Hamiltonian density
$\hat{H_i}(x)=\hat{H}_{em}(x)=\hat{J}_{\mu}(x)\hat{A}_{\mu}(x)$.
$\hat{A}_{\mu}(x)$ is taken in Feynman gauge with the indefinite metric.
The commutator of $\hat{H}_i(x),\hat{H}_i(x')$ is c-value function
for the regarded c-value currents $J_{\mu}$,
 which permits to transform T-product into
the product of the integrals over 4-space :
\begin {equation} 
     \hat{S}_{em}(J)=exp[i\phi(J)-i\int\hat{H}_i(x)d^4x]=
 exp[(i-1)V(J)+U(J)]
\end {equation}
where
\begin {equation} 
 U(J)=i\sum_{\lambda=1,2}\int d\tilde{k}[J_{\mu}(\vec{k})e^{\lambda}_{\mu}
a^+(\lambda,\vec{k}]
-J^*_{\mu}(\vec{k})e^{\lambda}_{\mu}a(\lambda,\vec{k})) ,
\end {equation}
$$
 V(J)=\frac{1}{2(2\pi)^3}\int d\tilde{k}J_{\mu}^*(\vec{k})J_{\mu}(\vec{k})
$$
where $d\tilde{k}=\frac{d^3k}{k_0}$ ,$a(\lambda,\vec{k})$ is the 
photon annihilation
 operator (Friedrichs,1953).
Below we'll omit the sum over the photon polarization indexes
 $\lambda$ or the polarization  vectors $e^{\lambda}_{\mu}$,
  where it's unimportant.
 $\phi(J)=V(J)$ is equal to the quantum  phase between in- and out- states,
 if the relation $J^*_{\mu}(k)=J_{\mu}(-k)$  is fulfilled,  which is true
for  $J_{\mu}$ of (1).
As easy to see from (3) the amplitudes of
the production of the photons with the different momentum $\vec{k}$
are independent.
If the initial e-m field state is vacuum $|\gamma_0\rangle=|0\rangle$,
 then  the average number of the produced photons is
$ d\bar{N}_{\vec{k}}=c|J_{\mu}(\vec{k})|^2d\tilde{k}$.
The action of $\hat{S}_{em}(J)$   
  results in the divergent photon spectra 
$ d\bar{N}_{\gamma}=c\frac{dk_o}{k_0}$, for $J_{\mu}(k)$
of (1). It means that the final asymptotic
state $|f\rangle$ includes the infinite number of very soft photons
which total energy is finite
(Jauch,1954).
 In the same time  it gives :  
$$
|\langle f|0\rangle|=exp[-V(J)]=exp(-\frac{\bar{N_{\gamma}}}{2})
$$
It follows that the state $|f\rangle$ doesn't belong
 to initial photon Fock space
 $H_F$, but to the different Hilbert space orthogonal to $H_F$.
  So the complete field states manifold $M_c$ becomes nonseparable, i.e.
 described by the tensor
 product of the
infinitely many Hilbert spaces $H_i$, each of them having its own
cyclic vector - vacuum state $|0\rangle_i$.  
 Any state of $M_c$ is defined by
  two indexes $|\psi_j\rangle_i$, $i=0$ corresponds to $H_F$.    
 Remind that
any Hermitian operator $\hat{B}$ - observable transforms only vectors inside
 the same Hilbert
space $|\psi_2\rangle_i=\hat{B}|\psi_1\rangle_i$, 
and due to it for the arbitrary
$|\psi_1\rangle_i , |\psi_2\rangle_l$ , $i\neq l$, 
 $\langle_i\psi_1|\hat{B}|\psi_2\rangle_l=0$.
So if the final state is the superposition of the states 
from  different spaces $|f\rangle=
|f_1\rangle_i+|f_2\rangle_l$
the interference terms (IT) for  any $\hat{B}$ 
between $|f_1\rangle_i,|f_2\rangle_l$ are
 equal to zero. Consequently any measurement performed on
such  disjoint states 
 can't distinguish  the mixed and the pure initial states,
which permit to resolve  mentioned EOP for UN representations. 
 Note that  the 
  bremsstrahlung due to the charge classical motion results in
the final e-m field state which can belong only to the
 single Hilbert space $H_i$.  
Consequently to obtain the final disjoint states
described QED formalism must be extended
to incorporate the bremsstrahlung of
 the charged particles states superpositions, which will be done
in this paper.

The transition from $H_F$ to some $H_i$
 corresponds to Bogolubov boson transformation
 of the  free field operators $a(\lambda.\vec{k}),a^+(\lambda,\vec{k})$
\begin {equation} 
     b_(\lambda,\vec{k})=a(\lambda,\vec{k})+iJ_{\mu}(\vec{k})e^{\lambda}_{\mu}
\end {equation}
 which is nonunitary for $J_{\mu}(\vec{k})$ of (1), but conserves the vector
 norm  $\langle f|f\rangle=\langle 0|0\rangle$.

\section { QED Measurement Model}  
 As the example of the collapse induced by the bremsstrahlung
 we'll regard  Weak scattering
of the electron on the neutral particle (neutrino) $\nu$ with mass $m_0$,
which in principle can be zero.
 We can consider also the electron coulomb scattering on the nucleus
 ,but its infinite range results in the infinite 
total cross section which introduce the unnecessary complications.
We'll show that the final photon bremsstrahlung disjoint states formally
 measure whether the act of scattering took place or
the particles passed unscattered and conserved their initial state. In the same
 time it's the measurement of the $e$  helicity $\lambda_e$, because  for 
its left, right helicities  cross-sections 
  $\sigma_L>>\sigma_R$ in weak interactions.


Now  $e$ motion is nonclassical and defined by $e$ field operators
  the general S-operator for
 $\hat{H}_i(x)=\hat{H}_{em}(x)+\hat{H}_w(x)$ should be found.  
Here we'll describe the method of its matrix elements  
$\langle f|\hat{S}|i\rangle$ calculations for the  states of
 interest without
finding S-operator analytical form, which is quite difficult. This
 nonperturbative  calculations are possible for the soft photon radiation  
 for which BRF condition is fulfilled, i.e. the total e-m field recoil is
much less then $e$ momentum transferred to $\nu$, as was discussed in chap.1. 
  It means that $\hat{H}_{em}(x)$
doesn't act on $e$ field operators conserving its spin and momentum
and acts only on e-m field operators (Jauch,1954).
On the contrary $\hat{H}_w$ acts only on $e,\nu$ fields, and due to it
it's possible to factorise S-operator into $\hat{S}_w$ and $\hat{S}_{em}$
parts. $\hat{S}_w$ defines the  skeleton diagram which
defines solely the final $e',\nu'$ states,
 which is dressed by the soft radiation given by $\hat{S}_{em}$. In its turn
$\hat{S}_{em}$ and consequently the
final radiation field depends on final  $e$ momentum, defined
 by $\hat{S_w}$  action on the initial state.  
 
So we should start from the calculation of $\hat{S}_w$ action on the
initial $e,\nu$ states neglecting $\hat{H}_{em}(x)$. 
 The smallness of the weak interaction constant $G$ permits
to calculate  $\hat{S_w}$ perturbatevely with the good accuracy,
 and 
  at c.m.s. energies below 1 TeV , which we'll regard here,
 its calculation
 can  be restricted to  1st order diagram (Cheng,1984). Its
 amplitude $M_w$ for the weak vertex $e,\nu\rightarrow e',\nu'$
 results in the  spherically
symmetric distribution of $e'$,$\nu'$ in the c.m.s  :
\begin {equation} 
   M_w(e',\nu')=\langle e,\nu|\hat{S}^1_w|e',\nu'\rangle=
\frac{G}{\sqrt{2}}J_{L\mu}J^{*}_{L\mu}=\bar{u'}_e\gamma_{\mu}
  (1+\gamma_5)u_{e}\bar{u'_{\nu}}\gamma_{\mu}(1+\gamma_5)u_{\nu}.
\end {equation}
 From it we can find the final e-m field state 
if we know the operator $\hat{S}_{em}(J^l)$ for the 
initial and final momentum eigenstates $|e\rangle , |e'_l\rangle$. 
Despite that now  $e$ electromagnetic current is formally the operator,
 it was found that  $\hat{S}_{em}(J^l)$ is independent of the initial and
final $e$ polarisations and
 described by the formulae (2)  
 in which current Fourier transform is equal to
 $J^l_{\mu}=J_{\mu}(k,p,p'_l)$ of (1), where $p,p'_l$ are
the corresponding eigenvalues (Jauch,1954). This result doesn't seems surprising
, because such states describe the prescribed $e$ motion
in the phase space.
Then, as follows from  the superposition principle,
 if $e$ final momentum eigenstate
 $|e'_l\rangle$ have the amplitude $c_l$ ,the final
system state is :
$$
    |\psi_f\rangle=
\sum c_l |e'_l\rangle|\nu'_l\rangle\hat{S}_{em}(J^l)
|\gamma_0\rangle      
$$
  In our case
it results in the final nonclassical system state which is 
the entangled product of $e'$,$\nu'$ states and disjoint e-m field states
\begin {equation} 
    |f_w\rangle=\sum_{l=0}c_l|f\rangle_l= |f\rangle_{\alpha}+|f\rangle_0=
\sum_{l=1} M_w(e'_l,\nu'_l) |e'_l\rangle|\nu'_l\rangle
|\gamma^f\rangle_l+M_0|e\rangle|\nu\rangle|0\rangle      
\end {equation}
where $|\gamma^f\rangle_l=\hat{S}_{em}(J^l)|0\rangle$,  the sum over $l$ means
the integral over the correlated final $e',\nu'$ momentum $p'_l,p'_{l\nu}$.
 $M_0$ is the zero angle amplitude of particles nonscattering.
All  the partial phases $\phi_l$ are infinite ,
moreover , as follows from (3) their differences
$\delta_{lm}$ are divergent, as must be  for the disjoint states :
$$
 \delta_{lm}=\int\frac{d\tilde{k}[J^{l*}_{\mu}(\vec{k})J^l_{\mu}(\vec{k})-
J^{m*}_{\mu}(\vec{k})J^m_{\mu}(\vec{k})]}{2(2\pi)^3}
=F(\varphi_{lm})\int\frac{dk_0}{k_o}
$$
,where $\varphi_{lm}$ is the angle between $\vec{p'}_l,\vec{p'}_m$.
Due to it in the limit $t= \infty$ this process is formally 
completely irreversible
,because  
T-reflection of the sum of such states with the indefinite relative phases 
produces the new state completely different from the initial one.       

 Then ,as was stressed already, for such disjoint state
 any measurement of arbitrary Hermitan  $\hat{B}$  will give :
 $ \langle f_0|\hat{B}|f\rangle_{\alpha}=0$.
It means that we obtained in QED based model the final disjoint state
with the finite total energy. Its components $|f\rangle_0$ and
 $|f\rangle_{\alpha}$ corresponds to the different values of
$e$ polarisation $\lambda_e$ which we intended to measure. 
As the result this state have all the observable properties
 of the mixed state which must appear after this measurement.
Note that it was obtained for the complete final state without
averaging over some subsystem ,or regarding it  as the
unmeasurable environment (Zurek,1982).  
Formally this is the main result of our paper, yet
it's important to discuss also the practical aspects 
of the continuous photons spectra measurements, and possible developments
of QFT models for the real solid state detectors.

In practice $\hat{B}$ can be only
 the bounded operator in $H_F$
,because only this case corresponds to the photon measurements
by the finite detectors ensemble (Itzykson,1980).
Note that the practical direct  IT observation is impossible
even between the single photon $|\vec{k}\rangle$
and the vacuum states,as  follows from Photocounting theory 
  (Glauber,1963). It shows that all e-m field operators
 $\hat{B}_{\gamma}$ which can be measured are the functions of
 $\hat{n}(\lambda,\vec{k})=a(\lambda,\vec{k})a^+(\lambda,\vec{k})$ - the
photon number operators. But for such operators
 $ \langle\vec{k}|\hat{B}_{\gamma}|0\rangle=0$ , and the same will be
true for any states with the different photon numbers. 
 To reveal IT presence for this single photon state the only 
 possibility is to perform
the special premeasurement procedure (PP), namely 
 $|\vec{k}\rangle$
must be reabsorbed by its source $Q_{\gamma}$ and the interference of the
source states for some new observable of the form 
 $\hat{B}_s=a(\vec{k})\hat{B}$ studied. Yet, to our knowledge
 there is no general proof
that such PP always exists for the
multiphoton states with the continuous spectra.  
The famous Recurrence theorem is true only for the discrete spectra
(Bocchiery,1957).

 Such PP certainly doesn't exist for $|f_{w}\rangle$
states at $t=\infty$ due to discussed loss of relative phases between
its parts $|f\rangle_l$. Really if the phase differences $\delta_{lm}$
are infinite for the sum of e-m field states ,then their reabsorbtion
will mean that this loss of coherence is transferred
 to $Q_{\gamma}$ state which
 after it will become
 disjoint.  But we'll give the qualitative arguments
 that such PP probably doesn't exist
 also for this states taken  at finite time.  

As the example we'll regard  PP layout in which the scattered 
 $e,\nu$ are reflected by some very distant mirrors back to the 
interaction region where they  can rescatter again. Then we must calculate  
$e$ radiation appearing  after three consequent collisions, taking into
account also 'internal' $e$ radiation between collisions.
As  Low theorem demonstrates 
    e-m radiation field in the infrared limit
in any process is defined solely by the current calculated between
asymptotic in-,out- momentum eigenstates, neglecting
intermediate states (Low,1958). It means that we can
 apply for the calculations the method described above and in particular
the resulting formula (6). 

  Then the initial e-m field state restoration is defined by
  $ \langle 0|\hat{S}_{em}(J^l)|0\rangle$
amplitude of $|0\rangle$ restoration in the $e,\nu$ rescattering ,which
is nonzero  only for $J_{\mu}(k)=0$ as follows from (2).
It means that $e$ in- and out- momentum must coincide,and from the energy
conservation  the same be true for $\nu$. So we must calculate
the probability $P$ of 2-nd order weak process $i\rightarrow v'_l
\rightarrow v^r_l\rightarrow i$. Here $v'_l$ are all possible intermediate
 states and $v^r_l$ are  reflections of $v'_l$ 
in the nondispersive mirrors ,the reflection amplitude is supposed to be
  $M_r=exp(i\phi_c)$ and can be omitted. The calculation
is simplified by the spherical symmetry of weak scattering  and
we obtain, omitting some unessential details :
$$   
  P=\frac{\int|M_w(e'_l,\nu'_l)M_w(e^r_l,\nu^r_l\rightarrow e,\nu)|^2do_v}
{\int\int|M_w(e'_l,\nu'_l)M_w(e^r_l,\nu^r_l\rightarrow e_f,\nu_f)|^2do_vdo_f}
=\frac{|\bar{M}_w|^4o_v\delta^3(\vec{p_f}-\vec{p_e})}
{ |\bar{M'}_w|^4o_vo_f}=0
$$
   $o_v,o_f$ are the phase spaces of intermediate and final $e,\nu$ states
 which are reduced to the corresponding $e$ phase spaces. So $o_v,o_f$
is isomorphic to the spherical surface with the radius $r=|\vec{p}_e|$ with
the nearly constant density of the final states on it. 
$\bar{M}_w,\bar{M'}_w$ are the expectation values of $M_w$ over the indicated
phase volumes ,which are assumed to be of the same order.
   The restoration of the
initial state corresponds to a single point $\vec{r}_0=\vec{p_e}$
  on this surface. Each
 infinitely close point to $\vec{r}_o$ corresponds to another Hilbert space
 with the infinite number of soft photons. So the zero probability
of the initial state restoration obtains the simple geometrical interpretation
;  $\vec{r}_0$ is the 
single nonsingular point in the phase space which  can be omitted
without changing any physical result. Despite  this arguments
have qualitative character they demonstrate that
 the disjoint states evolution irreversibility is connected with the
QM principal uncertainty of scattering angles.

It's important to note that such effect supposedly can exist also for the
rescattering of the photon states with the finite NDF and continuous spectra
 which  belong to $H_F$. If the proof of it will be given,
 on which we work now,
 then the conditions of the collapse observations in QED can become
more tight and wouldn't demand the use of UN representations and disjoint
states. 

In QFT studies the situations in which the particular dynamics makes
some operators unobservable are well known. The most famous example
is QCD colour confinement where coloured
charge is the analog of electric charge and  QCD
Hamiltonian contains infrared
singularity induced by the massless bosons - gluons
 (Itzykson ,1980).  Any attempt 
  to measure coloured operators ,for example quark or
 gluon momentum, results in the soft gluon production.
 In a  very short time this coloured quanta fuses
  into some number of colourless hadrons and consequently only the hadron
 operators are the real observables of this theory.

In practice the measurement is performed on the localized states 
(wave packets) and lasts only the finite time. 
Meanwhile it was shown that any localized charged state 
includes the unlimited number of soft photons  (Buchholz,1991).
 It supposes that the structure
of localized and nonlocalized states asymptotically coincides
and their evolution will result in analogous disjoint 
final states. 

The real detectors are the localized solid objects to which the regarded model 
can't be applied directly. Yet QFT methods were used very successfully
in  Solid State Physics , and so we can hope that they'll permit to 
describe the collapse in the real detectors.  The example of such 
approach gives the simple model
of the collapse  induced  by   Ferromagnetic phase transition
 (Mayburov,1995).

It's well known that the  solid
 state collective excitations -  
quasiparticles are massless
and their excitation spectra have no gap i.e. infrared divergent
 (Umezawa,1982).
 This quanta readily interact with e-m field,
so any excitation of the crystal in the vacuum is to be relaxated by the
 soft radiation. Its main mechanism  is probably
the cascade phonon decay $P\rightarrow P'+\gamma$.  
So the excitation of the crystal by the measured energetic particle
can result in a new disjoint state of the crystal plus
the external electromagnetic field.  
 This idea can also be applicable for the finite
system if its surface is regular and
 can be regarded as the topological defect with 
infinite NDF  which results in
a special kind of boson condensation in the crystal volume (Umezawa,1978). 
   
In conclusion
 we've shown that final states of $e-\nu$ scattering asymptotically
in the standard S-matrix limit
reveal the properties of the mixed state i.e. perform the collapse.
This doesn't seems a surprise ,because the classical features of
electron bremsstrahlung states were stressed often (Kibble,1968). 
In addition this model  can formally describes the
radiation decoherence process of the
special kind ,when the system being measured generates
 its environment from the initial vacuum.


\begin {thebibliography}{99}

\bibitem {Ah} A.Akhiezer, V.Berestetsky 'Quantum Electrodynamics'
 (Nauka,Moscow,1981)

\bibitem {Boc} P.Bocchieri, A.Loinger Phys. Rev. 107,337 (1957)

\bibitem {Buch} D.Buchholz,M.Porrman,U.Stein Phys. Lett. B267,377 (1991) 

\bibitem {Che} T.P.Cheng,L.F.Li 'Gauge theory of Elementary Particles'
(Claredon,Oxford,1984)

\bibitem {Desp} W. D'Espagnat, Found Phys. 20,1157,(1990)

\bibitem {Fri} K.O.Friedrichs 'Mathematical Aspects of The
Quantum Theory' (Interscience Pub.,N-Y,1953)

\bibitem {Fuk} R. Fukuda Phys. Rev. A ,35,8,(1987)

\bibitem {Gla} J.R.Glauber Phys. Rev. 131,2766,(1963)

\bibitem {Its} C.Itzykson,J.Zuber 'Quantum Field Theory'
 (McGraw-Hill,N.Y.,1980)

\bibitem {JA} J.M.Jauch, F.Rohrlich, Helv. Phys. Acta 27, 613, (1954)

\bibitem {Kib} T.Kibble Phys. Rev. 175,1624 (1968), J. Math. Phys. 9,315 
 (1968) 
\bibitem {Lo} F.Low Phys. Rev. 110,974 (1958)

\bibitem {May} S.Mayburov Int. Journ. Theor. Phys. 34,1587(1995)

\bibitem {Nee} Y.Neeman ,Found. Phys ,16,361   (1986)

\bibitem {Ume} H.Umezawa,H.Matsumoto, M.Tachiki Thermofield 
Dynamics and Condensed States (North-Holland,Amsterdam,1982)

\bibitem {Umu} H.Umezawa et. al. Phys.Rev B18,4077 (1978)

\bibitem {Zur} W.Zurek, Phys Rev, D26,1862 (1982)
\end {thebibliography}
\end{document}